\documentclass[a4paper,11pt]{article}

\usepackage{jinstpub} 
\usepackage{graphicx}
\title{Development of a Negative Ion Micro TPC Detector with SF$_{6}$ Gas for the Directional Dark Matter Search}

\author[a,b]{T. Ikeda,\note{Corresponding author.}}
\author[a]{T. Shimada,}
\author[a]{H. Ishiura,}
\author[a]{K. D. Nakamura,}
\author[a]{T. Nakamura,}
\author[a] {K. Miuchi}

\affiliation[a]{Department of Physics, Graduate School of Science, Kobe University, Kobe, Hyogo, 657-8501, Japan}
\affiliation[b]{Division of Physics and Astronomy, Graduate School of Science, Kyoto University, Kyoto, 606-8502, Japan}

\emailAdd{ikeda.tomonori.2s@kyoto-u.ac.jp}

\abstract{
A negative ion micro time projection chamber (NI$\mu$TPC) was developed and its performance studied.
An NI$\mu$TPC 
is a novel technology that enables the measurement of absolute $z$ coordinates for self-triggering TPCs. 
This technology provides full-fiducialization analysis, which is not possible with conventional gaseous TPCs, and is useful for directional dark matter searches in terms of background rejection and the improvement of the angular resolution.
The developed NI$\mu$TPC prototype had a detection volume of 12.8 $\times$ 25.6 $\times$ 144~mm$^{3}$.
The absolute $z$ coordinate was determined with a location accuracy of 16~mm using minority carrieres of SF$_{5}^{-}$.
Simultaneously, there was a successful reconstruction of the three-dimensional (3D) tracks  with a spatial resolution of 130~$\mu\rm{m}$.
This is the first demonstration of 3D tracking with the detection of absolute $z$ coordinates, and it is an important step in improving the sensitivity of directional dark matter searches.
}


\arxivnumber{} 

\begin{document}
\maketitle
\flushbottom


\section{Introduction}
Dark matter in the universe remains one of the unsolved mysteries in physics. Despite many worldwide experimental efforts, no experiment has reached a widely agreed discovery of the dark matter. Directional dark matter searches are said to provide clear evidence of the direct detection. Gaseous time projection chambers (TPCs) have been studied as a directional detector because they can detect the tracks of recoiled nuclei. Several groups, such as DRIFT~\cite{2017APh....91...65B}, NEWAGE~\cite{doi:10.1093/ptep/ptv041}, and MIMAC~\cite{1742-6596-1029-1-012005}, have developed gaseous TPCs, and measurements have been performed in underground laboratories.

For rare-event search experiments like dark matter searches, fiducialization is a powerful tool for removing external and internal backgrounds. 
An example of the successful application of this analytical technique is a two-phased detector using liquid noble gas~\cite{PhysRevLett.119.181301}. 
In this application, the time difference between the first (before the drift) and the second (after the drift) signals is used to reconstruct the absolute $z$ coordinate. 
Here, the $z$ coordinate is defined as the drift direction.
Unfortunately, the absolute $z$ coordinate cannot be reconstructed with conventional gaseous TPCs because there is no effective way to know the time of the event as the scintillation signal was used in the two-phased liquid noble gas detectors.
Therefore it was thus considered impossible to achieve the full fiducialization with gaseous TPCs. 
However, a discovery of ``minority carriers'' in a CS$_{2}$ $+$ O$_{2}$ gas mixture by the DRIFT group enabled to measure absolute $z$ coordinates and broadened the potential for the gaseous TPCs~\cite{BATTAT20151}. 
In electro-negative gases like $\rm CS_2$, electrons are captured by the molecules shortly after the interaction, and the negative ions instead of electrons are drifted. If more than two species of negative ions are produced, the  difference in velocity can be used to measure the absolute $z$ coordinate.
Following the discovery of the minority carriers in the $\rm CS_2 + O_2$ gas mixture,  SF$_{6}$, which is a safer gas compared to $\rm CS_2$, was found to perform in a similar way~\cite{1748-0221-12-02-P02012}.
In addition, fluorine has a large cross-section for spin-dependent interaction. Therefore, $\rm SF_6$ is considered to have excellent properties as a TPC gas for the directional dark matter search. Several studies on $\rm SF_6$ gas have been conducted in recent years~\cite{Ikeda,Baracchini_2018,TomPhD}.

Despite the importance of 3D tracking performance as a directional detector, there have been no studies on 3D tracking for 
$\rm SF_6$-based TPCs.
This is simply  due to the lack of micro-patterned gaseous detectors (MPGDs) coupled 
with readout electronics suitable for negative ion TPCs. 
We developed a prototype negative ion micro TPC (NI$\mu$TPC) with a micro-patterned gaseous detectors (MPGDs) and originally-developed electronics.  
This paper describes a study of NI$\mu$TPC performance, including the first demonstration of 3D trackings with absolute $z$ coordinate reconstruction.

\section{NI$\mu$TPC detector}
\label{sec:apparatus}
In this section, the experimental measurement setup is described for the TPC, the readout electronics, and the operating system. 

\subsection{Negative ion micro TPC}
\begin{figure}[h]
    \centering
    \includegraphics[width=12cm]{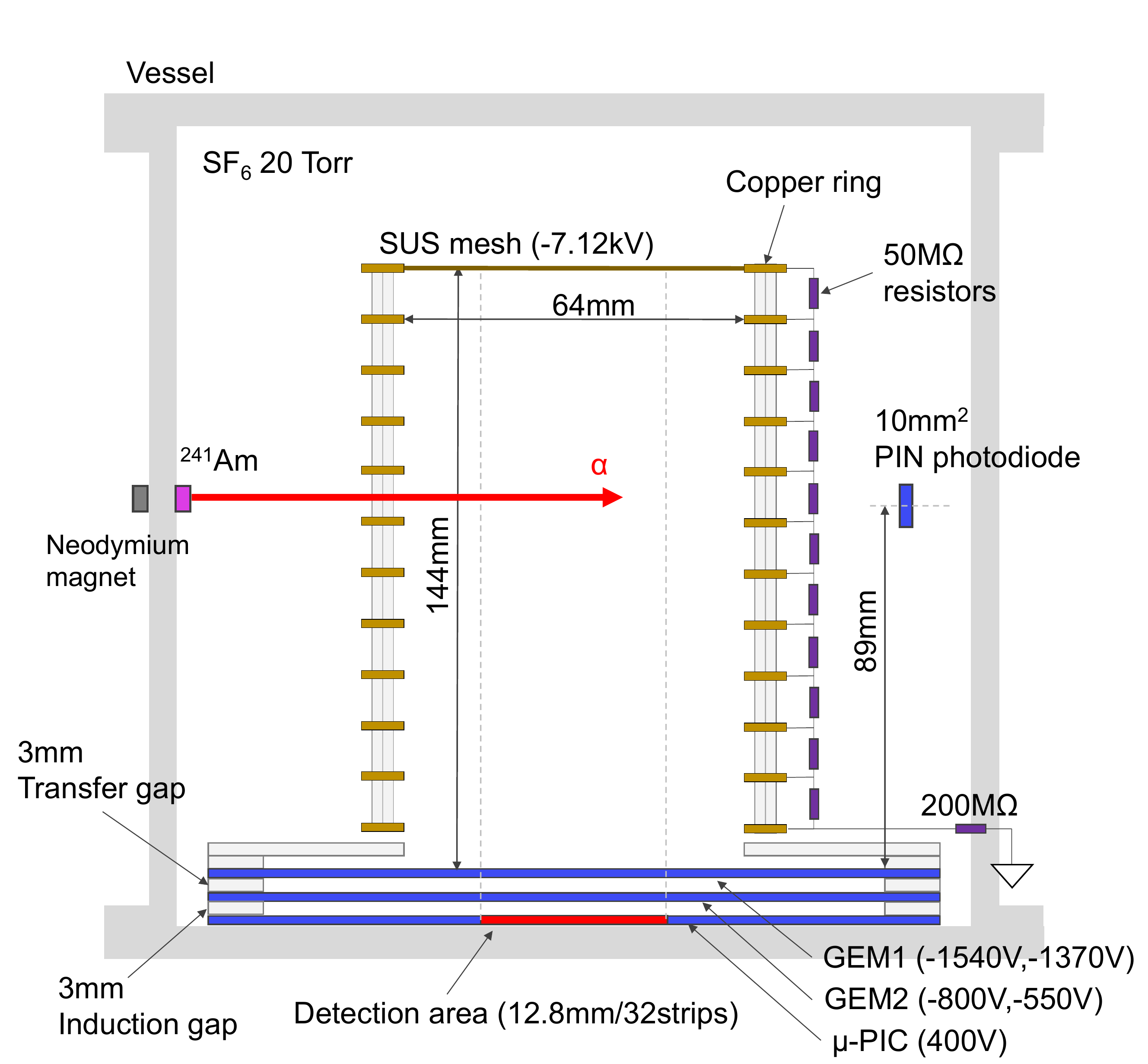}
    \caption{Schematic structure of the NI$\mu$TPC. 
             The detection volume was formed by 12 copper rings connected with 50 M$\Omega$ resistors. 
             The gas amplification section consisted of two GEMs and the $\mu$-PIC. 
             The PIN photodiode was mounted at a drift length of 89 mm to measure drift velocity.
             A $^{241}$Am source was set and controlled by a neodymium magnet. }
    \label{fig:schematic_structure}
\end{figure}

A schematic drawing of the NI$\mu$TPC is shown in Figure~\ref{fig:schematic_structure}. 
The gas amplification section consists of two layers of gas electron multipliers (GEMs; SciEnergy Co., Ltd.) and a micro pixel chamber ($\mu$-PIC; Dai Nippon Printing Co., Ltd.) ~\cite{2007NIMPA.573..195T}. 
These devices were arranged in a cascade with a 3~mm transfer gap and a 3~mm induction gap.
The substrate of the GEM was a 100~${\mu}$m thick liquid crystal polymer with a hole size and pitch of 70~${\mu}$m and 140~${\mu}$m, respectively. 
The  $\mu$-PIC had 256~$\times$~256 pixels with a pitch of 400~$\mu$m, which was read with 256 anode and 256 cathode strips. The anode and cathode strips were orthogonally formed, and two-dimensional imaging are realized by taking the coincidence of these strips. 
The cathode strips had circular openings with a diameter of 250~$\mu$m.
The anode strips were formed on the backside of the substrate.
Cylindrical anode electrodes with a diameter of 60~$\mu$m were formed on the anode strips, piercing the substrate at the center of the cathode openings (see Ref~\cite{2007NIMPA.573..195T} for details). 
The detector was operated at a gas gain of 1,900, with the bias voltages shown in Figure~\ref{fig:schematic_structure}.
The effective areas of gas multiplication for both GEM and $\mu$-PIC were 10~$\times$~10~cm$^{2}$; however, the readout area was limited to 1.28 $\times$ 2.56~$\rm cm^2$ due to the number of readout channels (32 + 64 strips).

A TPC detection volume with a drift length of 144~mm was formed using
12 copper rings, each with an inner diameter of 64~mm. 
These rings were connected using 50~M$\Omega$ resistors. 
The drift plane was made of stainless-steel mesh, and a
negative voltage of $-7.12$~kV was supplied, forming an electric field of 0.40~kV/cm in the detection volume.
A $^{241}$Am $\alpha$-ray source was set inside the vessel; the effective size of the source was 4~mm in diameter.
The source position was controlled from outside the vessel using a neodymium magnet. 

The performance of negative ion SF$^{-}_{6}$ TPC is known to be affected by water vapor due to out-gassing \cite{1748-0221-12-02-P02012}. Water vapor contamination was monitored with a  dew point meter (DMT152, Vaisala). 
At the beginning of each measurement, to reduce
the initial water contamination,
the vessel was evacuated to below 1~Pa, then flushed and filled with SF$_{6}$ gas at 20 Torr.
 A gas circulation system with Zeolum$^{\textregistered}$ (A-3) was installed to capture the water vapor produced by the out-gassing. 
The water contamination level was maintained below 300~ppm during the measurement. 

\subsection{Data acquisition system}
\begin{figure}[h]
    \centering
    \includegraphics[width=12cm]{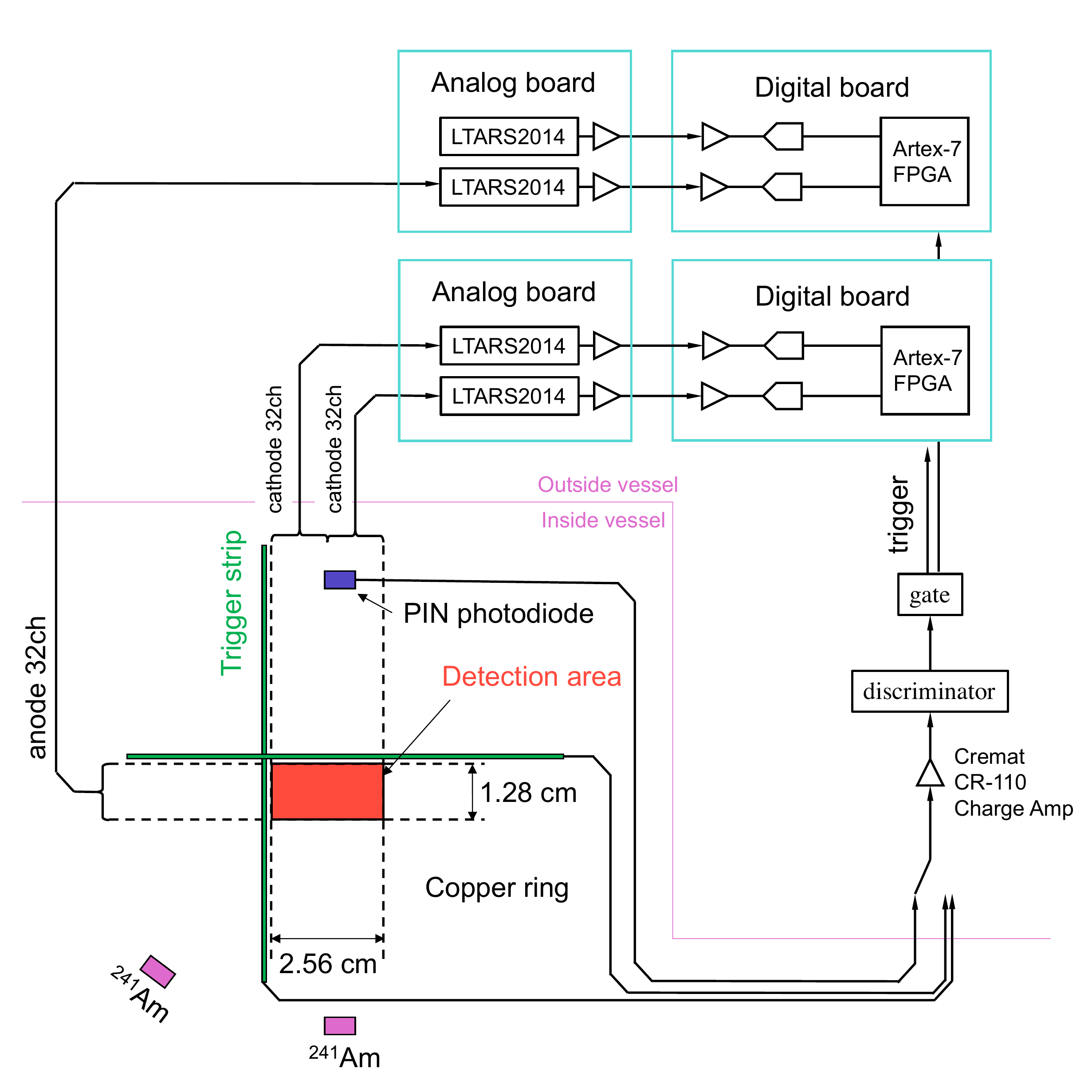}
    \caption{Schematic of the data acquisition system.
             The red region is the detection area.
             The $\mu$-PIC anode signals (32 channels) and cathode signals (64 channels) were processed by LTARS2014 chips. 
             The output signals were digitized and sent to a computer.
             The triggers were made by either the PIN photodiode or the $\mu$-PIC cathode or anode signal.}
    \label{fig:DAQ}
\end{figure}

 A schematics of the data acquisition system is shown in Figure~\ref{fig:DAQ}. 
 The 32 anode-strips and 64 cathode-strips were read by  preamplifier chips (LTARS2014~\cite{LTARS2014}) through 100~pF capacitors.
 The preamplifier chip was a low-noise ASIC developed by KEK group in Japan for liquid argon TPC detectors. 
 The amplified waveformes were digitized by a digital board at a sampling rate of 2.5~MHz.
 The dynamic range and resolution of the digitization were 2~V and 12~bit, respectively. 
 The specifications of the analog and digital boards are summarized in Table~\ref{tab:readout}.
 The data stored in the memory of the digital board were sent to the computer using SiTCP technology~\cite{SiTCP}, as required by the trigger signal. 
 The trigger was made one of two ways: a PIN-photodiode trigger for the measurement of drift velocity or a $\mu$-PIC trigger created by the signal from the cathode or anode strip next to the detection area for 
3D tracking measurements.

\begin{table}[htb]
	\begin{center}
	\caption{Parameters of readout electronics}
		\begin{tabular}{l l} \hline
		Analog board parameters &  \\ \hline
		Conversion gain & 10 mV$/$fC \\ 
		Shaping time & 1  $\mu$s \\
		Channel number & 2 chips $\times$ 32~ch/chip  \\ 
		ENC & 6,000 at 100 pF \\ \hline
		Digital board parameters &   \\ \hline
		Sampling rate & 2.5 MHz \\
		Depth of sampling & 4,000 \\
		ADC dynamic range & 2 V \\
		ADC & 12~bits \\ \hline
		\end{tabular}
		\label{tab:readout}
	\end{center}
\end{table}

\section{Performance of the NI$\mu$TPC}
\label{sec:NImTPC}
The performance of the NI$\mu$TPC, the detection efficiency of minority carriers, the location accuracy of the absolute $z$ coordinate, and the spatial resolution of the 3D reconstruction are described in this section.

\subsection{Minority carrier detection}
\label{sec:minority_detection}
\begin{figure}
    \centering
    \includegraphics[width=\linewidth]{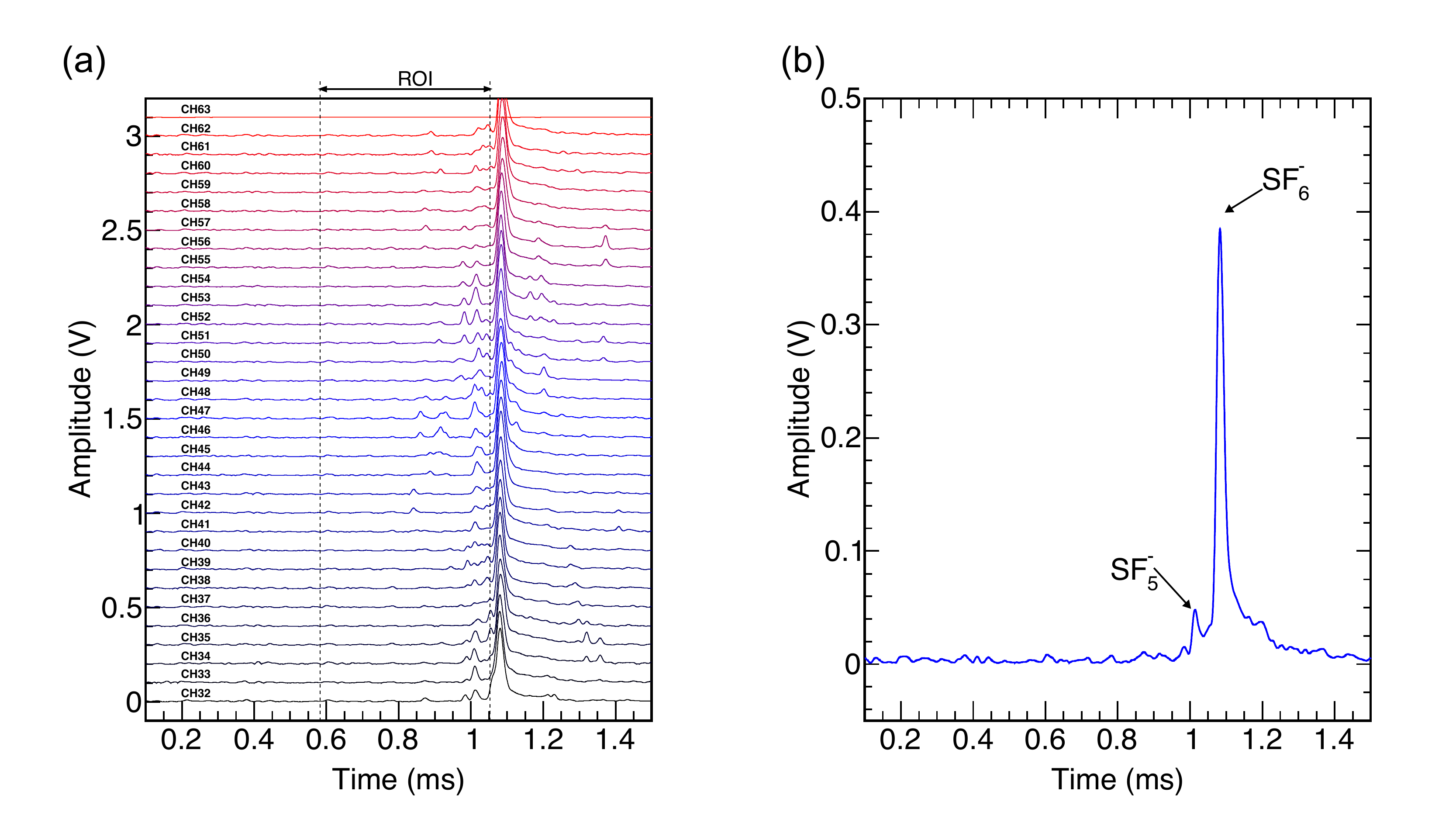}
    \caption{(a) Example of 32 channel signals from $\mu$-PIC anode strips demonstrating the minority carrier signatures. The position of the major peaks corresponds to the drift time of 89 mm. Each waveform is shown with an offset of 100 mV per channel number for clear viewing. (b) Averaged waveform of 32 channel anode signals. Two peaks are clearly visible in the waveform. The major peak and the minority peak were created by SF$^{-}_{6}$ and  SF$^{-}_{5}$, respectively. }
    \label{fig:waveform}
\end{figure}

Measurement was performed with the $^{241}$Am source set at a position of $z=89$~mm. This measurement was used to confirm the detection of minority carriers generated at a known $z$ coordinate. 
The $\alpha$-rays were emitted from the source, which passed through the detection volume and were detected by the PIN photodiode located on the opposite side.
The signal from the PIN photodiode was used to determine the event timing, ${\it i.e.}$, the time-zero of the drift time. 
The waveforms were smoothed with a Gaussian filter to suppress high-frequency noise during the first stage of  analysis. 
Examples of 32-anode signals and an averaged waveform are shown in Figures~\ref{fig:waveform}a and \ref{fig:waveform}b, respectively.
A major peak due to SF$_{6}^{-}$ negative ions and a minority peak due to SF$_{5}^{-}$ can be seen in both waveforms and are more clearly visible in the averaged waveform. As the time-zero was set by the PIN photodiode, the signal time indicated that the drift time of these ions corresponded to a drift length of 89~mm.
Average times for major peaks were calculated from thousands of events, and a drift velocity of $8.1\pm0.2~\rm{cm/ms}$ was obtained.
Minority carrier drift velocity ($8.9\pm0.2~\rm{cm/ms}$) was obtained by the same method.
These drift velocities were used to determine the absolute $z$ coordinate, as discussed below.

It should be noted that some waveforms show more than two peaks in Figure~\ref{fig:waveform}a.
Figure~\ref{fig:waveform}b shows a wide range components between the minority peak and the main peak.
The waveform structure depends on the chemical reactions associated with electron capture in SF$_{6}$ gas.
In particular, as explained in Ref~\cite{1748-0221-12-02-P02012}, water vapor contamination and the strength of the electric field in the drift region have a significant effect on the production of negative ions.
In this study, the main source of water vapor was out-gassing from acrylic plates that were placed to prevent discharge from the copper rings. H$_{2}$O contamination in SF$_{6}$ gas creates stable SF$_{6}^{-}$(H$_{2}$O)$_{n}$ clusters and can produce  negative ions SOF$_{4}^{-}$ and F$^{-}$(HF)$_{2}$ as final products. 
These negative ions create fake peaks at similar time as SF$_{5}^{-}$, which reduces the accuracy of absolute $z$ determination.
Previous studies~\cite{1748-0221-12-02-P02012} have shown that these effects are suppressed by a high electric field of approximately 1,000 V/cm$^{2}$. This should possible with our future detectors.

\subsection{Detection efficiency of minority carriers}
\label{sec:efficiency}
\begin{figure}
\centering
\includegraphics[width=7cm,clip]{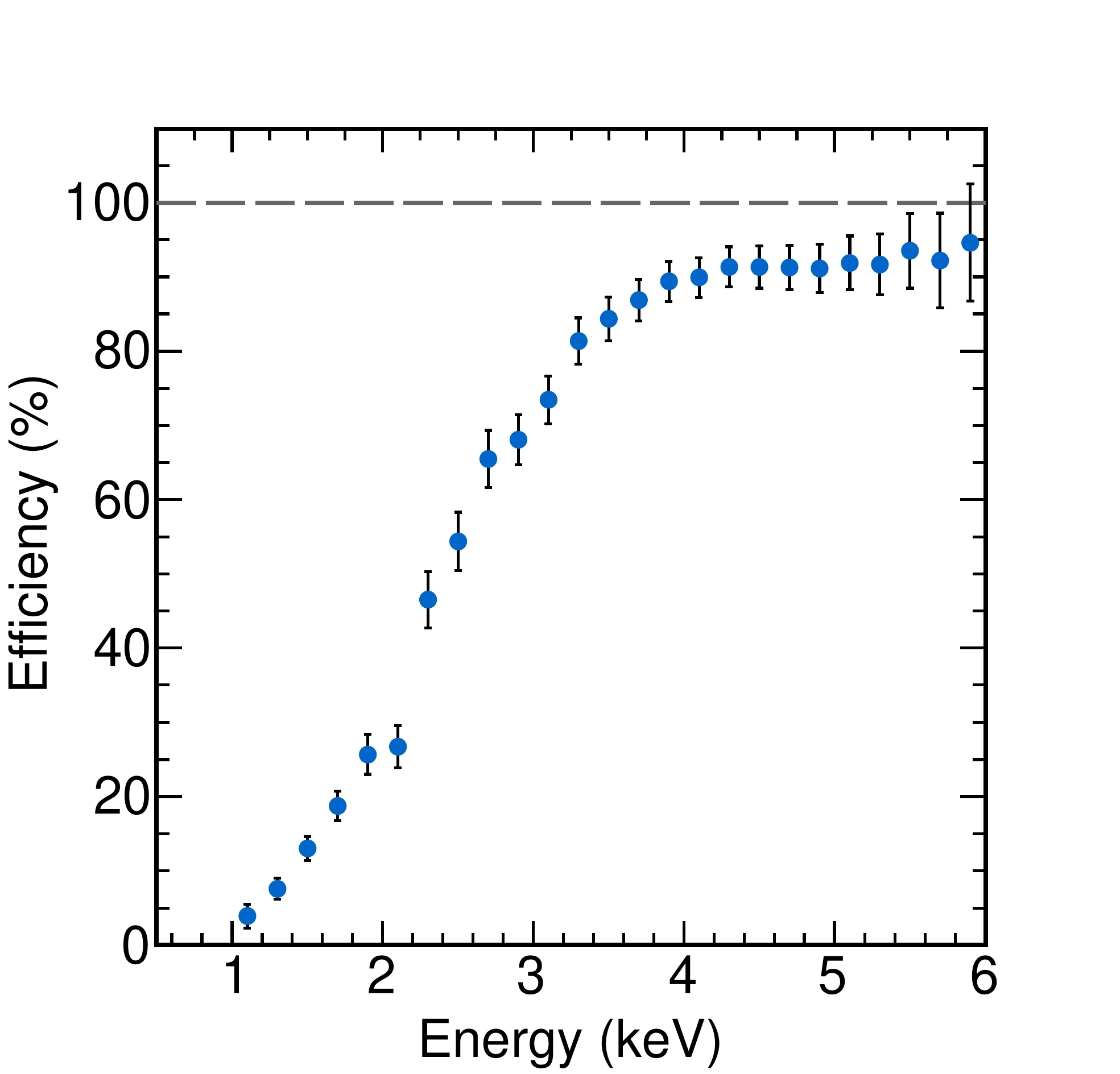}
\caption{Detection efficiency of the minority peak as a function of the energy deposition to one strip.}
\label{minority_eff}       
\end{figure}

Minority carrier detection efficiencies were evaluated using the same data used in previous studies.
The region of interest (ROI) time for the minority-peak search was set at 
between $30$~$\mu$s and $700$~$\mu$s prior to the main peak timing ($-700~\mu s < \rm{Time} < -30~\mu s)$.
Peaks were searched in the ROI of each anode strip waveform using an analytical threshold.
Figure~\ref{minority_eff} shows the detection efficiency of minority carriers as a function of energy deposition in one strip.
Detection efficiency increases as energy deposition increases, reaching a plateau of $\sim$90$\%$ at $\sim$4 keV.
Here, the energy deposition of the $\alpha$-rays on each strip is known by the linear energy transfer calculated in the SRIM simulation~\cite{SRIM}.
Tracking the fluorine nuclei of O(10~keV) is a major consideration in the dark matter search applications.
The energy deposition of a fluorine nuclei of 10 keV in SF$_{6}$ gas 20 Torr has been found by the SRIM simulation to be approximately 3.4 keV per strip (400 $\mu$m).
Therefore, this NI$\mu$TPC was found to perform sufficiently well in terms of detection efficiency in minority peaks of fluorine nuclei. 

\subsection{Determination of absolute $z$ coordinates}
\label{sec:z_calib}
The difference in the arrival times of SF$^{-}_{5}$ and SF$^{-}_{6}$ can be used to reconstruct the absolute $z$ coordinate in practical dark matter detector applications.
A measurement of the $z$ coordinate reconstruction was performed using this method.
The $z$ coordinate was determined by the following equation:
\begin{equation}
z= \frac{v_{\rm m} \cdot v_{\rm M}}{v_{\rm m} -v_{\rm M}} \Delta T
\end{equation}
where $v_{\rm m}$ and $v_{\rm M}$ are the drift velocities of the minor negative ion (SF$^{-}_{5}$) and major negative ion (SF$^{-}_{6}$), respectively.

\begin{figure}
\centering
\includegraphics[width=7cm,clip]{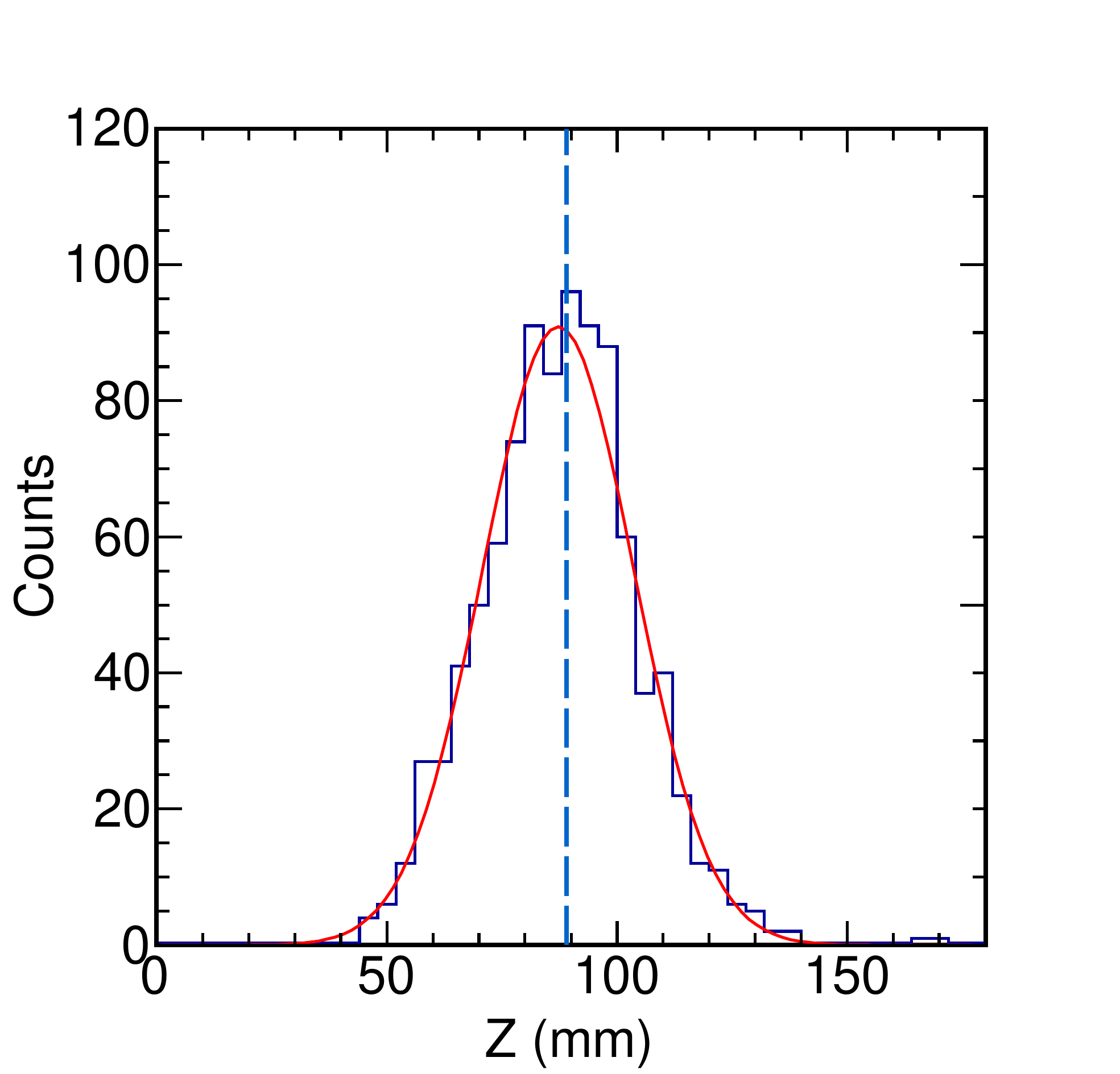}
\caption{Reconstracted absolute $z$ coordinate distribution. The $^{241}$Am source position is shown as the blue dashed line.}
\label{fig:absoluteZ}    
\end{figure}

To assess the accuracy of the absolute $z$ determination, the $^{241}$Am was set at $z=89$~mm, and the data were acquired using the PIN photodiode trigger.
The peak finding algorithm used in previous studies was  applied for the anode 32 strips, and the averaged time difference was set as $\Delta T$. 
The reconstructed absolute $z$ coordinates of thousands of  events are shown in Figure~\ref{fig:absoluteZ}, together with the $^{241}$Am source position.
The difference between the actual source position and the mean value of the distribution was 1.2 mm; therefore, the reconstructed $z$ coordinate was in good agreement with the source position. The location accuracy of 16 mm for one event was obtained as the $\sigma$ of the Gaussian fit.
As mentioned in Section~\ref{sec:minority_detection}, the fake peaks, which are created by negative ions of other species, worsened the location accuracy and confused the peak finding algorithm.

\begin{figure}
\centering
\includegraphics[width=7cm,clip]{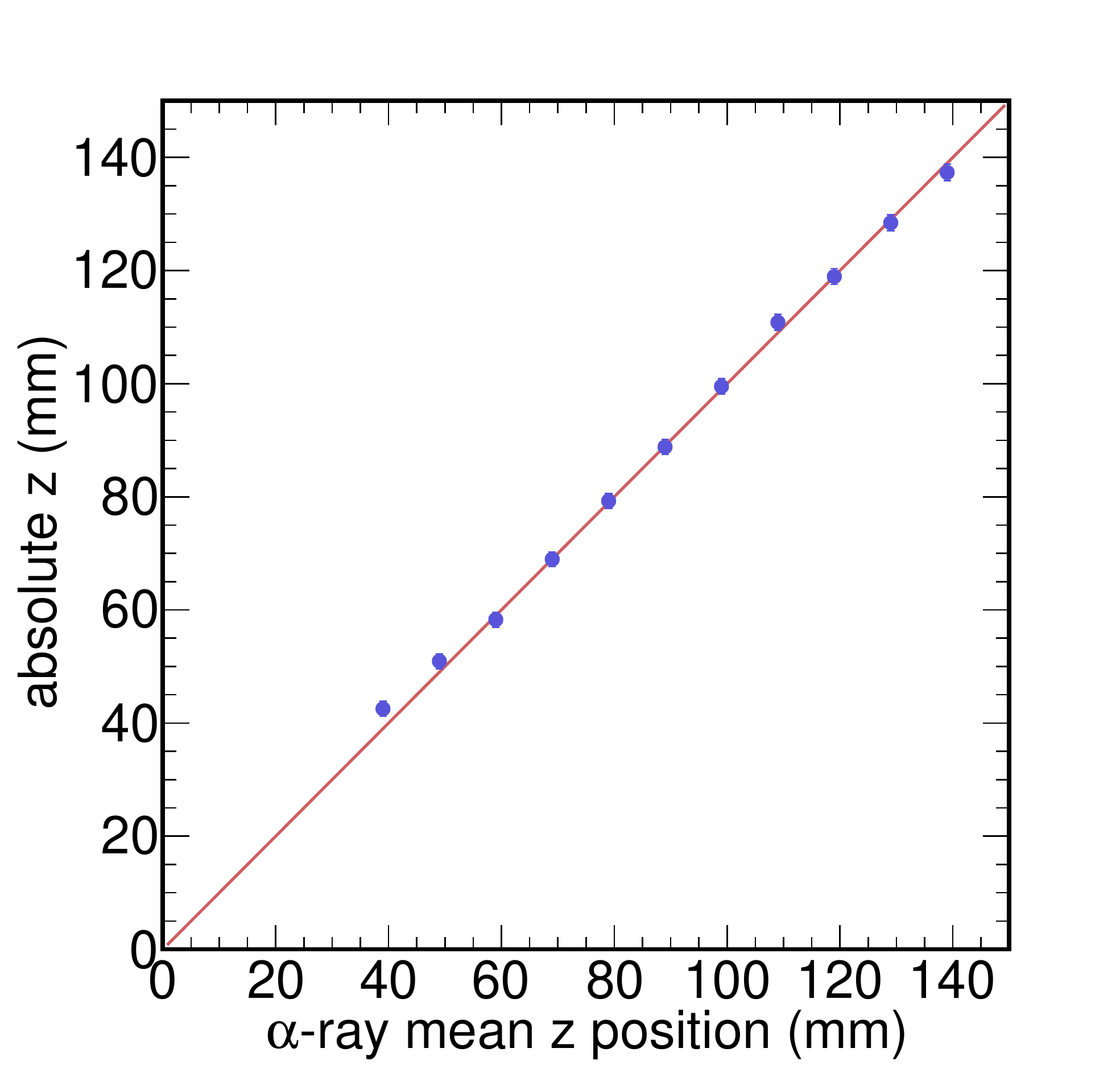}
\caption{Reconstructed absolute $z$ coordinate for each source position.
The red line indicates the reference function for $y=x$.}
\label{fig:Z_calib}     
\end{figure}

The $^{241}$Am position was then scanned from 39~mm to 139~mm at 10~mm intervals, and the absolute $z$ coordinate  was determined using the same method. 
The trigger from the $\mu$-PIC anode strip was used in the same way as the dark matter detector. 
Tracks with an elevation angle $\theta$ of $-5^{\circ}<\theta<5^{\circ}$ were chosen to select the tracks parallel to the detection plane.
This selection was made to evaluated the intrinsic $z$ reconstruction without any deterioration caused by the widened $z$ coordinate.
Figure~\ref{fig:Z_calib} shows the mean value of reconstructed absolute $z$ distribution for each source position.
The error bars are the convolution of the systematic and statistical errors.
The main systematic error was the alignment of the source.
The red line is the reference line ($x=y$) for the actual position. As can be seen in Figure~\ref{fig:Z_calib}, the $z$ coordination was properly reconstructed within the error, and 
it can be concluded that the minority carrier can be used for the $z$ reconstruction of 39~mm~$<z<$~139~mm.
The peak finding algorithm did not work for events smaller than 39~mm due to the charge spread of the major peaks. 
This performance is expected to be improved in stronger electric fields.


\subsection{Reconstruction of three-dimensional tracks}
\label{sec:track}
\begin{figure}
\centering
\includegraphics[width=11cm,clip]{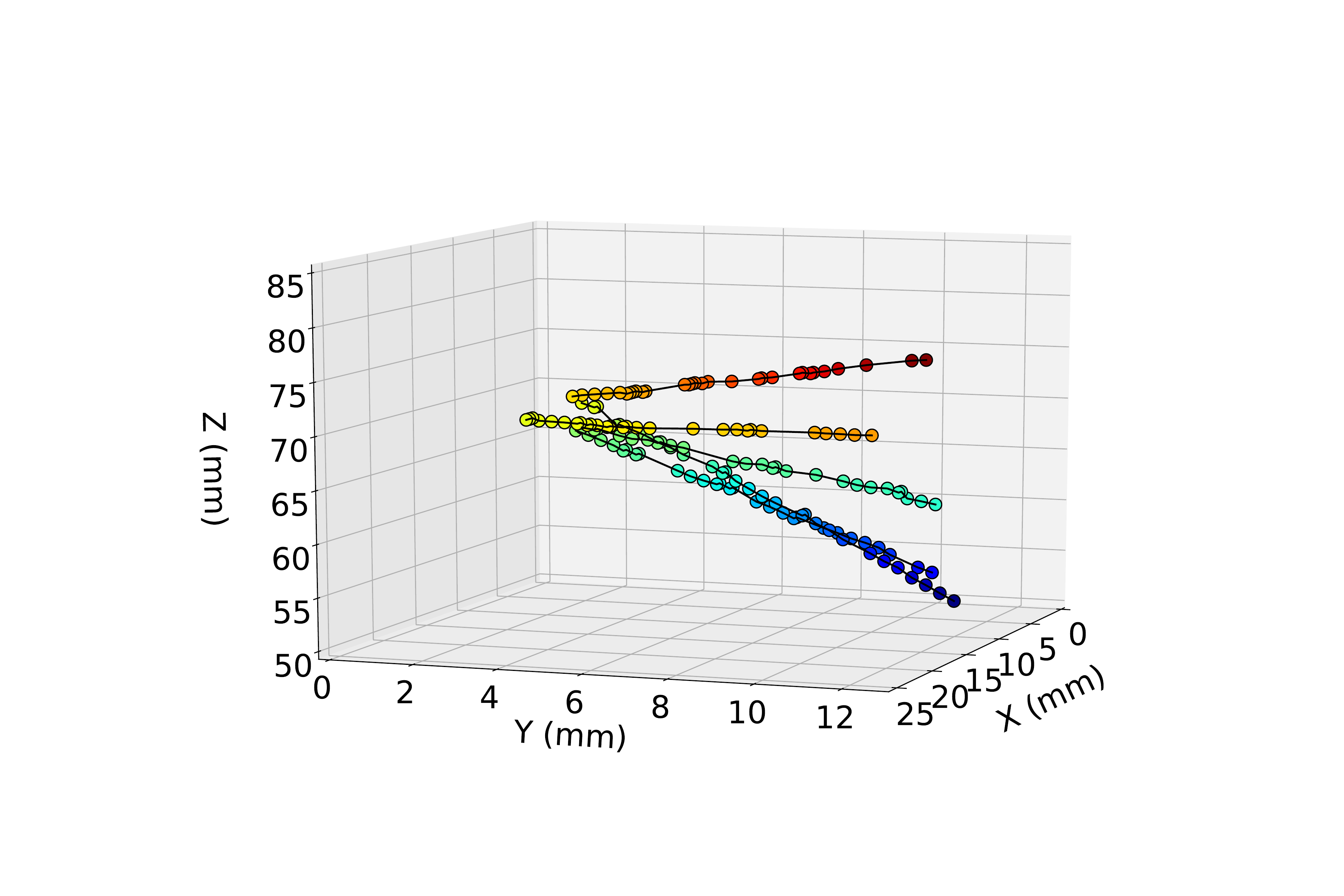}
\caption{Examples of 3D tracks reconstructed using the coincidence method.
The $z$ coordinate was determined by the minority carriers.}
\label{fig:track}    
\end{figure}

Finally, the reconstruction of 3D tracks was investigated.
The data were acquired by triggering the $\mu$-PIC cathode signal, and the time coincidence of the anode and cathode signals was also recorded.
The coincidence window was adjusted depending on the elevations.
Figure~\ref{fig:track} shows a typical examples of five reconstructed events.
The measured $z$ coordinate is the absolute $z$ coordinate, which was determined using the minority carriers.
It should be emphasized that this is the first demonstration to reconstruct absolute $z$-coordinates and 3D tracking simultaneously. 

\begin{figure}
\centering
\includegraphics[width=7cm,clip]{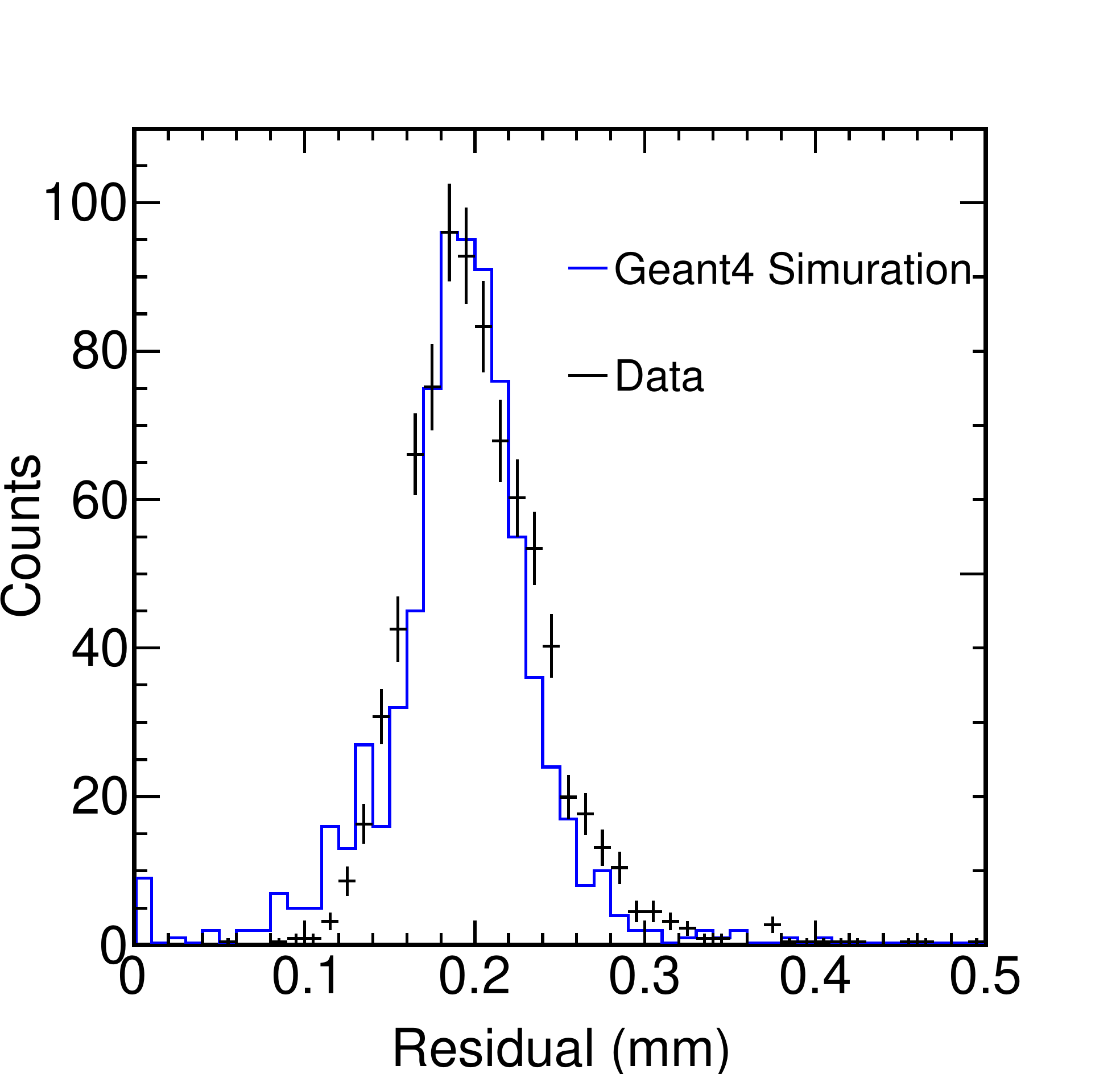}
\caption{Residual distribution of experimental data and Geant4 simulation. The simulation data was smeared with a 130~$\mu$m spatial resolution.}
\label{fig:3Dresolution}    
\end{figure}

To estimate the 3D spatial resolution for one anode-cathode coincidence or hit, we calculated the residual distribution of the hit and the fitted with a straight line. 
The events with an  elevation angle of smaller than 10$^{\circ}$ were  not used because the anode-cathode coincidence did not work.
The residual distribution was also calculated with a Geant4 simulation for several 3D spatial resolutions. Figure~\ref{fig:3Dresolution} shows the experimental data together with the simulation results at a 3D spatial resolution of 130 $\mu$m. 
This spatial resolution reproduced the experimental distribution best, and the value was comparable to the conventional electron-tracking micro TPC.
As as result, the NI$\mu$TPC was found to possess comparable tracking performance to the conventional TPC and additionally enable $z$ reconstruction or the full 3D fiducialization. Therefore, the NI$\mu$TPC is expected to expand the scope of directional dark matter searches.

\section{Conclusions}
A prototype NI$\mu$TPC with SF$_{6}$ gas was developed and its performance was studied.
The absolute $z$ coordinate was reconstructed with the minority peaks, obtaining a location accuracy of 16~mm for the absolute $z$ coordinate.
The first-ever reconstruction of simultaneous absolute $z$-determination and 3D tracking was demonstrated. 
A spatial resolution of 130 $\mu$m for one hit was obtained. 
These results indicate that the NI$\mu$TPCs can provide a similar tracking performance to conventional TPCs while allowing full 3D fiducialization. 
Therefore, the NI$\mu$TPC is expected to expand the reach of directional dark matter searches.


\acknowledgments
This work was supported by KAKENHI Grant-in-Aids (16H02189, 19684005, 23684014 and 26104005, 19H05806), JSPS Bilateral Collaborations (Joint Research Projects and Seminars) program, ICRR Joint-Usage, and Program for Advancing Strategic International Networks to Accelerate the Circulation of Talented Researches, JSPS, Japan (R2607).



\bibliography{mybibfile}

\end{document}